\documentclass[reprint,aps,superscriptaddress]{revtex4}
\usepackage{amsmath}
\usepackage{graphicx}
\usepackage{dcolumn}
\usepackage{bm}

\usepackage{amssymb}
\usepackage{dsfont}
\usepackage{url}
\usepackage{caption,subcaption}
\usepackage[colorlinks=true,allcolors=blue]{hyperref}

\def\pp{\mathbf{p}}

\def\lan{\left\langle}
\def\ran{\right\rangle}
\def\delpp{\Delta^{\alpha \beta} p_{\alpha}p_{\beta}}

\begin{document}
\title{Transport coefficients of quasi-particle models within a new relaxation time approximation of the Boltzmann equation}
\author{Gabriel S. Rocha}
\email{gabrielsr@id.uff.br}
\affiliation{Instituto de F\'{\i}sica, Universidade
Federal Fluminense, Niter\'{o}i, Rio de Janeiro,
Brazil}

\author{Maurício N. Ferreira}
\email{mnarciso@ifi.unicamp.br}
\affiliation{Instituto de Física Gleb Wataghin, Universidade Estadual de Campinas, Campinas, São Paulo, Brazil}

\author{Gabriel S. Denicol}
\email{gsdenicol@id.uff.br}
\affiliation{Instituto de F\'{\i}sica, Universidade
Federal Fluminense, Niter\'{o}i, Rio de Janeiro,
Brazil}

\author{Jorge Noronha}
\email{jn0508@illinois.edu}
\affiliation{Illinois Center for Advanced Studies of the Universe \& Department of Physics, University of Illinois at Urbana-Champaign, Urbana, IL 61801, USA}

\begin{abstract}
We investigate the transport properties of a kinetic theory model that is tuned to describe the thermodynamic properties of QCD at zero chemical potential using a new formulation of the relaxation time approximation. In contrast to previous approaches, the latter is constructed to preserve the fundamental properties of the collision term of the Boltzmann equation for any energy-dependence of the relaxation time. A novel choice of matching conditions is implemented to ensure that the background mean-field depends only on the temperature even when the system is out of equilibrium. We provide a consistent analysis of how the transport coefficients of relativistic Navier-Stokes theory vary with the energy dependence of the relaxation time. We also show that the entropy production of this theory is consistent with the second law of thermodynamics and verify that it is independent of the matching conditions employed. We used this fact to calculate the matching independent combination of transport coefficients. 
\end{abstract}

\maketitle

\section{Introduction}

Ultrarelativistic heavy-ion collisions allow us to systematically produce and study the properties of hot and dense deconfined matter, the quark-gluon plasma. In these experiments, it was found that the quark-gluon plasma displays striking transport properties, with a shear viscosity that is considerably smaller than extrapolations \cite{Csernai:2006zz} based on weak-coupling calculations \cite{Arnold:2000dr}. More recent phenomenological studies based on fluid-dynamical models were also able to estimate the bulk viscosity of the quark-gluon plasma \cite{everett2021multisystem,Nijs:2020roc,Parkkila:2021yha}. These analyses showed an opposite behavior for this transport coefficient, suggesting that it can be orders of magnitude larger than predictions derived in the weak-coupling limit \cite{Arnold:2006fz}. Understanding these novel transport features of the quark-gluon plasma from first principles is
 a very challenging task that is currently beyond the reach of \emph{ab initio} methods \cite{Meyer:2011gj}. Thus, this problem is often addressed using several effective approaches, such as the holographic correspondence \cite{Maldacena:1997re,Policastro:2001yc,Policastro:2002se,Kovtun:2004de,finazzo2015hydrodynamic,Rougemont:2015ona,Finazzo:2016mhm,Rougemont:2017tlu,Grefa:2022sav} and effective kinetic theory \cite{hosoya1985transport,
Jeon:1995zm,sasaki2009bulk,Romatschke:2011qp,chakraborty2011quasiparticle,Bluhm:2012,Alqahtani:2015qja}.

In this paper we investigate the transport properties of hot and dense matter using an effective kinetic description. We consider an extrapolation of the quasiparticle picture in which the masses of the excitations display a temperature dependence that is tuned to describe the equation of state (EOS) of QCD in a wide range of temperatures (which includes the deconfinement transition region) according to lattice QCD calculations \cite{Borsanyi:2010cj}. We stress that the quasi-particles in the model are not to be identified with either quarks or with gluons -- they are effective degrees of freedom at high temperatures tuned to describe QCD thermodynamics.

We further simplify the collision term of this effective kinetic theory using the novel relaxation time approximation proposed in Ref.~\cite{rocha:2021}. This new approximation preserves the fundamental properties of the collision that stem from microscopic conservation laws and is crucial when considering general matching conditions \cite{bemfica:18causality,rocha21-transient} or energy-dependent relaxation times (such as those used in QCD-inspired models \cite{dusling2010radiative}). One of the main goals of this paper is to determine how the relaxation time approximation of Ref.\  \cite{rocha:2021} can be generalized to describe a system with vanishing net-charge and a temperature-dependent particle mass. We then use this description to calculate the transport coefficients that emerge in the traditional Chapman-Enskog formalism \cite{deGroot:80relativistic}, here implemented for an arbitrary choice of local equilibrium state. 

The paper is organized as follows: Section \ref{sec:QPmodel} reviews the main features of the kinetic theory approach in the presence of a temperature-dependent particle mass. Section \ref{sec:rBEandhydro} defines the hydrodynamic variables in terms of the single-particle distribution function. We also present our relaxation-time Ansatz for the collision term, following Ref.~\cite{rocha:2021}. In Section \ref{sec:coeffs} we obtain transport coefficients within the Chapman-Enskog expansion and show how their behaviors change with different energy-dependencies of the relaxation time. Section \ref{sec:ent-prod} discusses how hydrodynamic frame invariant coefficients can be constructed using the entropy production computed from the Boltzmann equation with temperature-dependent masses. In Section \ref{sec:concl} we present our conclusions and outlook. Appendix \ref{sec:sum-coeffs} provides a detailed account of the computation of transport coefficients. \emph{Notation}: We use a mostly minus metric signature, $g_{\mu\nu}=(+,-,-,-)$, and natural units $\hbar=c=k_B=1$.

\section{quasiparticle model}
\label{sec:QPmodel}

The relativistic Boltzmann equation \cite{deGroot:80relativistic} is an integro-differential equation describing the  evolution of the single-particle momentum distribution function, $f_\pp$, of a gas. It can be understood as an effective theory that emerges from quantum field theories in the weak coupling limit \cite{Calzetta:1986cq,Jeon:1995zm,Arnold:2002zm,Berges:2005md} and is often applied to describe the pre-equilibrium stage of the hot and dense matter produced in heavy ion collisions \cite{Kurkela:2015qoa,Keegan:2016cpi,Kurkela:2018vqr,Romatschke:2011qp,Alqahtani:2015qja}.

In this work, we consider an effective relativistic Boltzmann equation for quasiparticles with a temperature-dependent mass, $M(T)$, which reads \cite{Jeon:1995zm,DEBBASCH20091818} 
\begin{equation}
\label{eq:BoltzmannQuasi}
p^\mu \partial_\mu f_\pp + \frac{1}{2}\partial_i M^2(T) \partial^i_{(\pp)} f_\pp = C\left[ f_\pp\right] ,
\end{equation}
where $C\left[f_\pp\right]$ is the collision kernel describing the interactions experienced by the particles. For instance, for a one-component gas of classical particles interacting only through elastic scattering, $C\left[ f_\pp \right]$ takes the form
\begin{equation}
C\left[f_\pp\right] \equiv \int dQ \ dQ^{\prime} \ dP^{\prime} W_{pp' \leftrightarrow qq'} (f_{\mathbf{q}}f_{\mathbf{q}'}  -  f_{\pp}f_{\pp'}) ,
\end{equation}
where $W_{pp' \leftrightarrow qq'}$ is the transition rate and we defined the integral measure
\begin{equation} 
\int dP = \int \frac{d^3\pp}{(2\pi)^3 E_\pp} ,
\end{equation}
with the particle energy being given by $E_\pp = \sqrt{ \pp^2 + M^2 }$. We also used the shorthand notation
\begin{equation}
\partial^i_{(\pp)} \equiv \frac{\partial}{\partial \pp_i} .
\end{equation}
We remark that the temperature that was introduced in the mass must be defined by imposing matching conditions \cite{deGroot:80relativistic}. This will be discussed later in this section. 



We now turn our attention to the conserved currents, which are essential ingredients for a  hydrodynamic description. For the sake of simplicity, we shall consider a locally neutral gas (an assumption often applied in fluid-dynamical simulations of ultra-relativistic heavy-ion collisions) and will disregard any net-charge 4-current. Thus, we only need to consider the continuity equations describing energy-momentum conservation,
\begin{equation}
\label{eq:Tmunucons}
\partial_\mu T^{\mu\nu} = 0.
\end{equation}
This equation is derived by multiplying the Boltzmann equation \eqref{eq:BoltzmannQuasi} by the 4-momentum, $p^\nu$, and integrating it with the integration measure $dP$. Next, we employ the fundamental property of the collision term, $\lan p^\nu C\left[ f_\pp \right] \ran = 0$ \cite{deGroot:80relativistic}, where we use the following notation for integrals over the distribution function  
\begin{equation}
\lan \ldots \ran \equiv \int dP (\ldots) f_\pp . \label{eq:avg_def}
\end{equation}
The energy-momentum tensor is then identified as  
\begin{equation}
\label{eq:Tmunudef}
T^{\mu\nu} \equiv \lan p^\mu p^\nu \ran  + g^{\mu\nu} B .
\end{equation}
We note that due to the spatial dependence of the mass (induced by the temperature), it is necessary to introduce a temperature-dependent \emph{background field}~\cite{Gorenstein:1995vm,Jeon:1995zm,Romatschke:2011qp,Alqahtani:2015qja}, $B$. This quantity must satisfy the following differential equation 
\begin{equation}
\label{eq:dB}
\partial_{\mu} B = - \frac{1}{2} \partial_{\mu} M^2 \lan 1 \ran ,
\end{equation}
due to the conservation law \eqref{eq:Tmunucons}. This allows one to determine $B$ dynamically, as shown in \cite{Alqahtani:2015qja}. We point out that in the presence of a conserved charge, the usual definition of the particle diffusion current, $N^\mu = \lan p^\mu \ran$, would not be modified by the introduction of the quasiparticle thermal mass~\cite{Romatschke:2011qp,Alqahtani:2015qja}.

A common feature shared by all methods for computing transport coefficients is an \emph{ad hoc} definition of a local equilibrium state, which serves as the starting point in any perturbative scheme that aims at describing the near-equilibrium behavior of many-body systems \cite{Israel:1979wp}. This choice, which defines the so-called hydrodynamic frame \cite{Israel:1979wp,Kovtun:2012rj}, affects the meaning of the standard hydrodynamic variables, such as temperature, flow velocity, and chemical potential, in an out-of-equilibrium state. The role played by the choice of hydrodynamic frame in establishing causality and stability of hydrodynamic theories was investigated in \cite{bemfica:18causality,Kovtun:2019hdm,bemfica:19nonlinear,Hoult:2020eho,bemfica:20general,Noronha:2021syv,Speranza:2021bxf}. In this work, we set our choice of hydrodynamic frame to simplify Eq.~\eqref{eq:dB}. We impose the following matching condition,
\begin{equation}
\label{eq:match1}
\int dP f_\pp \equiv \int dP f_{0 \pp},       
\end{equation}
with $f_{0\pp}$ being given by
\begin{equation}
\label{eq:EQL-qp}
f_{0\pp} \equiv \exp\left( - \beta u_{\mu}p^{\mu} \right).
\end{equation}
Expression \eqref{eq:match1} defines the temperature, $T = 1/\beta$. Another condition must be imposed in order to define the 4-velocity $u^\mu$ (which obeys $u_\mu u^\mu =1$). In this case, we impose the general constraint, 
\begin{equation}
\begin{aligned}
\label{eq:matching_kinetic1}
    \int dP \, q_{\textbf{p}}\, p^{\langle \mu \rangle} f_{\textbf{p}}  = \int dP \, q_{\textbf{p}}\, p^{\langle \mu \rangle} f_{0\textbf{p}},
\end{aligned}
\end{equation}
where we define the projection $p^{\lan \mu \ran} = \Delta^{\mu \nu}p_{\nu}$ with respect to the space-like projector $\Delta^{\mu\nu} = g^{\mu \nu} - u^{\mu} u^{\nu}$.
Since we only consider fluids at vanishing chemical potential, this condition will not be relevant for the results presented in this paper, which will not depend on the choice of the function $q_\pp$.

\subsection*{Temperature dependence of the quasiparticle mass}

Due to our choice of matching conditions, the field $B$ satisfies the following equation of motion
\begin{equation}
\label{eq:dB2}
\partial_{\mu} B = - \frac{1}{2} \partial_{\mu} M^2 \lan 1 \ran_0 ,
\end{equation}
where $\langle \cdots \rangle_{0}$ denotes momentum integrals over the local equilibrium distribution function. Thus, we can consider solutions of $B$ that depend only on temperature and can be determined solely by the quasiparticle mass. With this assumption, the field $B$ can be determined as if the system were in equilibrium, 
\begin{equation}
\label{eq:dBeq}
\frac{\partial B}{\partial T} = - \frac{1}{2} \frac{\partial M^2}{\partial T} \lan 1 \ran_0 .
\end{equation}
We note that more general methods to determine $B$ could have been considered. Nevertheless, the approach described above considerably simplifies the calculations needed in this paper. 

The equilibrium energy density, $\varepsilon_0$, and pressure, $P_0$, are given by   %
\begin{eqnarray}
\varepsilon_0 &=& \lan \left( u\cdot p \right)^2 \ran_0 + B(T), \label{eq:eps} \\ 
P_0 &=& - \frac{1}{3}\lan \Delta_{\mu\nu} p^\mu p^\nu \ran_0 - B(T), \label{eq:P} 
\end{eqnarray}
where the subscript `$0$' denotes local equilibrium. The entropy density, $s_0$, is obtained from the following thermodynamic relation,
\begin{equation}
\label{eq:thermo_consistency}
s_0 = \frac{\partial P_0}{\partial T} .
\end{equation}

We now discuss how $M(T)$ and $B(T)$ are determined for a given EoS (at zero chemical potential). To that end, we express the thermodynamic quantities in \eqref{eq:eps}, \eqref{eq:P}, and \eqref{eq:thermo_consistency} in terms of Bessel functions,
\begin{eqnarray}
\varepsilon_0 &=& \frac{gT^4w^2}{2\pi^2}[3K_2(w) + w K_1(w) ] + B(T) , \label{eq:eps_K} \\ 
P_0 &=& \frac{gT^4w^2}{2\pi^2}K_2(w) - B(T), \label{eq:P_K}\\
s_0 &=& \frac{g T^3 w^3 }{2\pi^2}K_3(w) , \label{eq:s_K}
\end{eqnarray}
where $g$ is the degeneracy factor, $w\equiv M(T)/T$, and $K_n(w)$ is the $n$--th modified Bessel function of the second kind \cite{gradshteyn2014table}. Thus, given an EoS, \emph{i.e.}, for a known $s_0(T)$, Eq.~\eqref{eq:s_K} can be solved to obtain the quasiparticle mass, $M(T)$. Moreover, since $wK_3(w)$ is a monotonically decreasing function of $w$, the solution $M(T)$ for a given $s_0(T)$ is unique. The field $B(T)$ can then be obtained from Eq.~\eqref{eq:dBeq}, which we rewrite as
\begin{equation}
\frac{\partial B(T)}{\partial T} = - \frac{g T M^2}{2\pi^2}K_1(w)\frac{\partial M(T)}{\partial T}.
\end{equation}
Integrating over $T$ and imposing the boundary condition~\cite{Romatschke:2011qp,Alqahtani:2015qja} $B(0) = 0$, gives
\begin{equation}
\label{eq:B_fin}
B(T) = - \frac{g}{2\pi^2}\int_0^T dx\, M^2(x) x\, K_1\left(\frac{M(x)}{x}\right) \frac{\partial M(x)}{\partial x} . 
\end{equation}

\begin{figure}[!h]
    \centering
\begin{subfigure}{0.47\textwidth}
  \includegraphics[width=\linewidth]{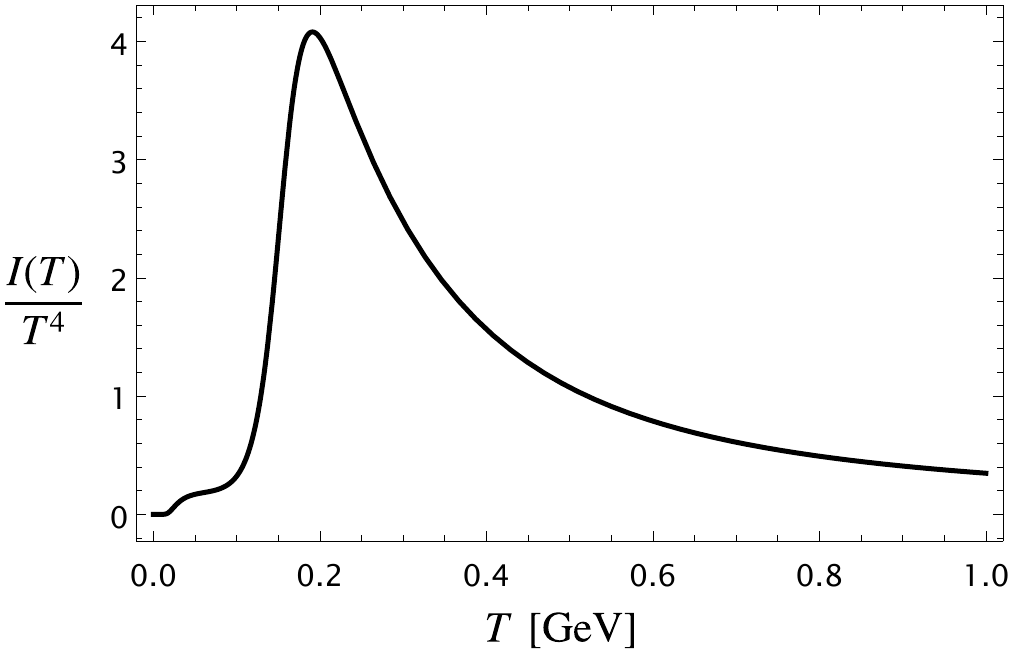}
\end{subfigure}\hfil 
\begin{subfigure}{0.47\textwidth}
  \includegraphics[width=\linewidth]{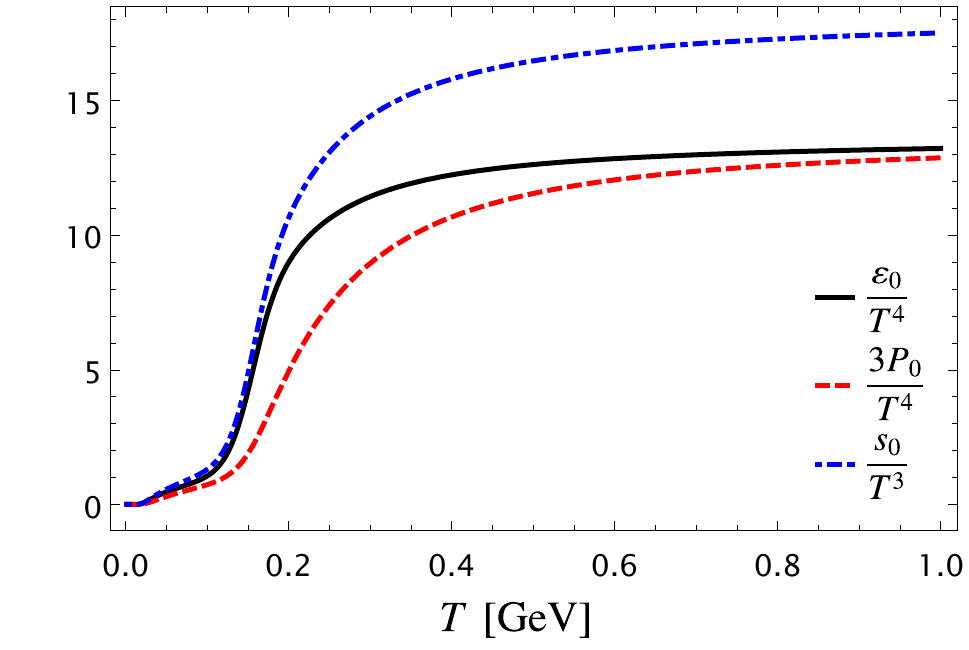}
\end{subfigure}\hfil 

\caption{Left: Dimensionless trace anomaly, $I(T)/T^4$, from the Wuppertal-Budapest Collaboration~\cite{Borsanyi:2010cj} lattice simulations, constructed using the parametrization given by Eq.~\eqref{eq:WP_tr_anom}. Right: Corresponding  energy density (black solid), thermodynamic pressure (red dashed), and entropy density (blue dot-dashed) of the quasiparticle model.}
\label{fig:1}
\end{figure}

\begin{figure}[!h]
    \centering
\begin{subfigure}{0.47\textwidth}
  \includegraphics[width=\linewidth]{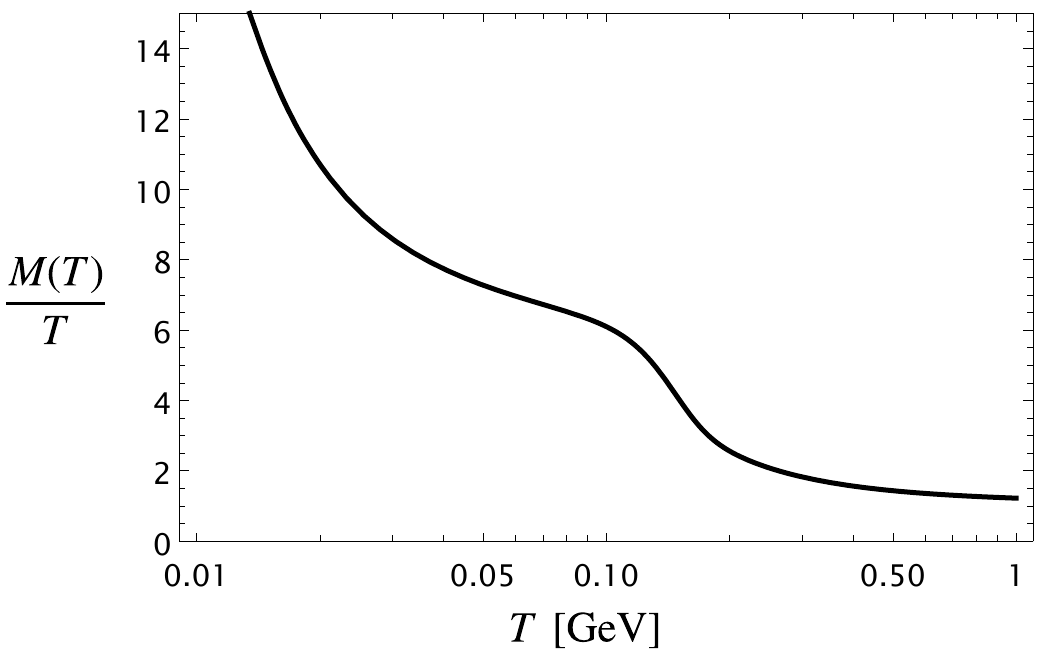}
\end{subfigure}\hfil 
\begin{subfigure}{0.47\textwidth}
  \includegraphics[width=\linewidth]{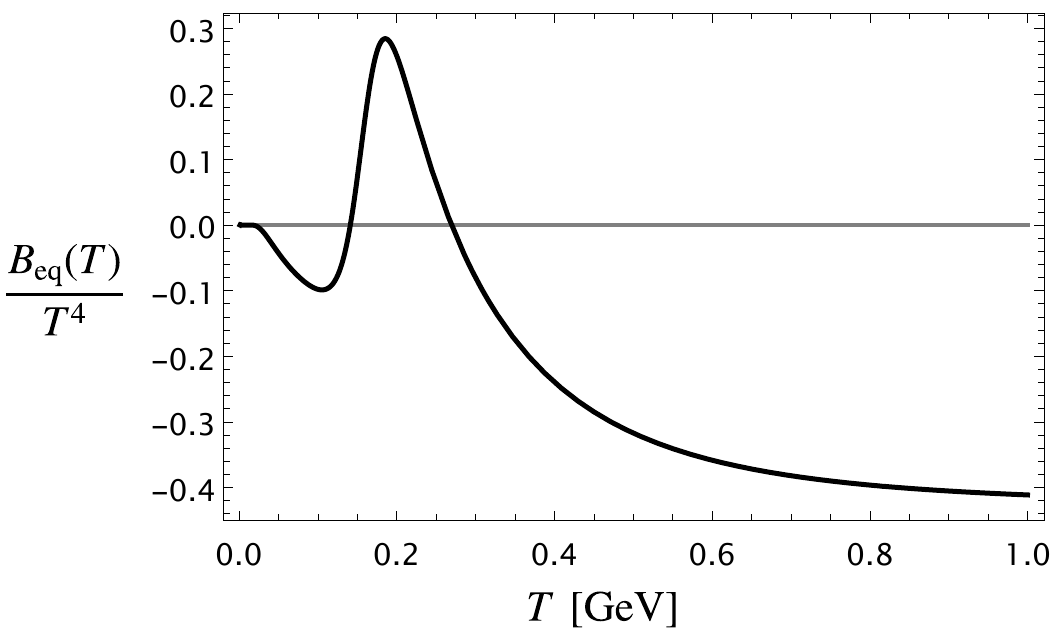}
\end{subfigure}\hfil 

\caption{Quasiparticle mass, $M(T)$, (left panel) and background field, $B_{\rm{eq}}(T)$, (right) which reproduce the EoS shown in Fig.~\ref{fig:1}. The mass is shown in logarithmic scale for the $x$ axis due to its fast evolution in $T$.}
\label{fig:2}
\end{figure}

\begin{figure}[!h]
    \centering
\begin{subfigure}{0.47\textwidth}
  \includegraphics[width=\linewidth]{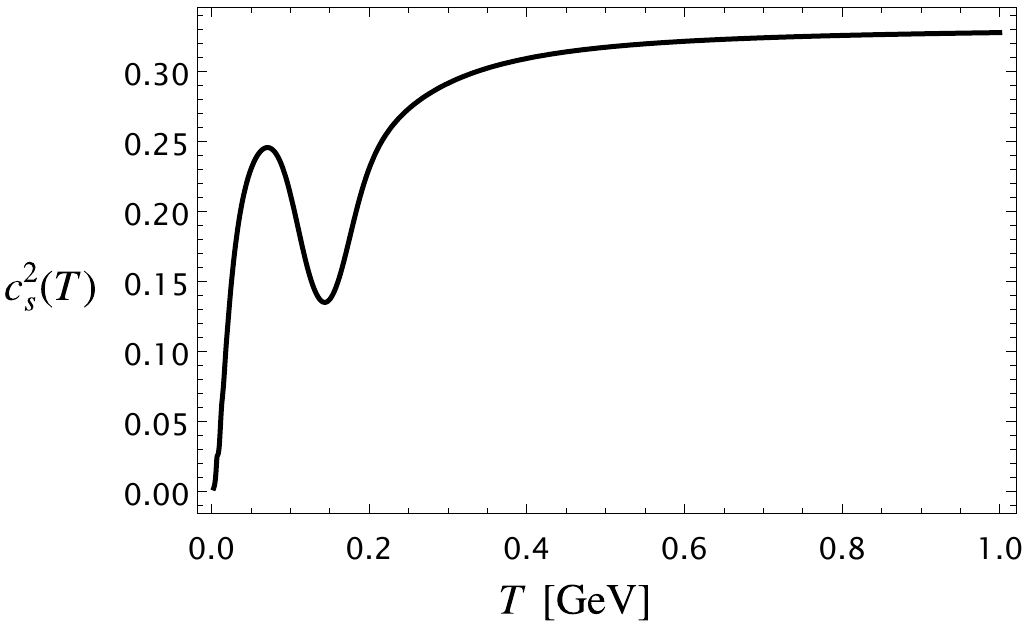}
\end{subfigure}

\caption{Speed of sound squared, $c_s^2$, as a function of temperature resulting from the EoS of Fig.~\ref{fig:1}.}
\label{fig:3}
\end{figure}

In this work, we apply the procedure described above using the lattice QCD EoS for $2 + 1$ quark flavors calculated by the Wuppertal-Budapest Collaboration~\cite{Borsanyi:2010cj}. The corresponding trace anomaly, $I \equiv \varepsilon_0 - 3 P_0$, has the analytic parametrization~\cite{Borsanyi:2010cj,Nopoush:2015yga,Alqahtani:2015qja}
\begin{equation}
\label{eq:WP_tr_anom}
\frac{I(T)}{T^4} = \left\lbrace \frac{h_0}{1 + h_3 t^2} + \frac{f_0\left[ \tanh(f_1 t + f_2 ) + 1 \right] }{1 + g_1 t + g_2 t^2} \right\rbrace\exp\left( - \frac{h_1}{t} - \frac{h_2}{t^2} \right) ,
\end{equation}
with parameters given by $h_0 = 0.1396$, $h_1 = - 0.1800$, $h_2 = 0.0350$, $f_0 = 2.76$, $f_1 = 6.79$, $f_2 = -5.29$, $g_1 = - 0.47$, $g_2 = 1.04$ and $h_3 = 0.01$, where $t \equiv T/( 0.2 \text{ GeV})$. The equilibrium pressure is related to the trace anomaly by the following thermodynamic relation \eqref{eq:thermo_consistency}
\begin{equation}
T\frac{\partial P_0(T)}{\partial T} - 4P_0(T) = I(T).
\end{equation}
Then, solving the above differential equation with the initial value $P(0) = 0$ yields
\begin{equation}
\frac{P_0(T)}{T^4} = \int_0^T \frac{dX}{X} \frac{I(X)}{X^4} \,, \label{eq:P_from_I}
\end{equation}
which determines also the energy density through $\varepsilon_0 = I + 3P_0$. Lastly, the thermodynamic relation, $\varepsilon_0 + P_0 = s_0 T$, allows the calculation of the entropy density. The trace anomaly, $I(T)$, resulting from \eqref{eq:WP_tr_anom} is shown in the left panel of Fig.~\ref{fig:1}, while the corresponding energy density, thermodynamic pressure, and entropy density are shown in the right panel of the same figure. From the obtained entropy density, the quasiparticle mass and background field are computed by means of Eqs.~\eqref{eq:s_K} and \eqref{eq:B_fin}, and we set $g$ to
\begin{equation}
g = \frac{\pi^4}{180}\left[ 4(N_c^2 - 1) + 7 N_c N_f \right] \,,
\end{equation}
with $N_c = N_f = 3$. The results are shown in Fig.~\ref{fig:2}. Finally, we show in Fig.~\ref{fig:3} the resulting speed of sound squared, $c_s^{2}$, as a function of temperature. At this point, we remind the reader that the quasiparticles in this model should not be mistaken by the degrees of freedom of QCD. Instead, they are effective degrees of freedom whose mass and degeneracy are tuned to reproduce lattice QCD thermodynamics.

\section{The relativistic Boltzmann equation and hydrodynamic variables}
\label{sec:rBEandhydro}

In general, the energy-momentum tensor can be decomposed in terms of the 4-velocity as   
\begin{equation}
\begin{aligned}
\label{eq:decompos-tmunu}
    T^{\mu \nu} &= \varepsilon u^{\mu} u^{\nu} - P \Delta^{\mu \nu} + h^{\mu} u^{\nu} + h^{\nu} u^{\mu} + \pi^{\mu \nu},
\end{aligned}
\end{equation}
where $\varepsilon$ is the total energy density, $P$ is the total isotropic pressure, $h^{\mu}$ is the energy diffusion 4-current, and $\pi^{\mu \nu}$ is the shear-stress tensor. These fields can then be expressed as integrals of the single-particle distribution function $f_{\mathbf{p}}$ using \eqref{eq:Tmunudef},
\begin{equation}
\begin{aligned}
\label{eq:def_kinetic}
     \varepsilon \equiv \lan E_{\textbf{p}}^2  \ran + B(T), \qquad P \equiv -\frac{1}{3} \lan \Delta_{\mu \nu}p^{\mu}p^{\nu} \ran - B(T),  \\
     h^{\mu} \equiv \lan E_{\textbf{p}} p^{\langle \mu \rangle} \ran, \qquad \pi^{\mu\nu} \equiv \lan p^{\langle \mu}p^{\nu \rangle} \ran, 
\end{aligned}
\end{equation}
where we defined the rank two irreducible projection $p^{\langle \mu} p^{\nu \rangle} = \Delta^{\mu \nu \alpha \beta} p_{\alpha} p_{\beta}$, which makes use of the double-symmetric and traceless tensor $\Delta^{\mu \nu \alpha \beta} = (1/2) \left( \Delta^{\mu \alpha} \Delta^{\nu \beta} + \Delta^{\mu \beta} \Delta^{\nu \alpha} \right) - (1/3) \Delta^{\mu \nu} \Delta^{\alpha \beta}$. In equilibrium, $h^{\mu}$ and $\pi^{\mu \nu}$ are identically zero and $\varepsilon = \varepsilon_{0}$ and $P=P_{0}$. Out of equilibrium, taking into account the matching conditions \eqref{eq:match1}, we have non-equilibrium corrections for the energy density and the pressure, so that $\varepsilon = \varepsilon_{0} + \delta \varepsilon$ and $P = P_{0} + \Pi$, where 
\begin{equation}
\begin{aligned}
\label{eq:def_kinetic2}
    \delta \varepsilon \equiv \left\langle E_{\textbf{p}}^2 \phi_{\textbf{p}} \right\rangle_{0}, \qquad 
    \Pi \equiv -\frac{1}{3} \left\langle \Delta_{\mu \nu}p^{\mu}p^{\nu}\phi_{\textbf{p}} \right\rangle_{0}, 
\end{aligned}
\end{equation}
and we defined
\begin{equation}
\begin{aligned}
\label{eq:phi}
\phi_{\bf p} \equiv \frac{f_{\bf p} - f_{0\bf p}}{f_{0\bf p}}.
\end{aligned}    
\end{equation}

In this setting, the equations of motion corresponding to the projection of Eq.~\eqref{eq:Tmunucons} on the fluid 4-vector $u^{\mu}$ and to its space-like projection are
\begin{subequations}
 \label{eq:basic-hydro-EoM}
\begin{align}
\label{eq:hydro-EoM-eps}
 D\varepsilon_{0}+D\delta \varepsilon + (\varepsilon_{0}+\delta \varepsilon + P_{0} + \Pi) \theta - \pi^{\mu \nu} \sigma_{\mu \nu} + \partial_{\mu}h^{\mu} + u_{\mu} D h^{\mu} &= 0, \\
\label{eq:hydro-EoM-umu}
(\varepsilon_{0} + \delta \varepsilon + P_{0} + \Pi)D u^{\mu} - \nabla^{\mu}(P_{0} + \Pi) + h^{\mu} \theta + h^{\alpha} \Delta^{\mu \nu} \partial_{\alpha}u_{\nu} +  \Delta^{\mu \nu} D h_{\nu} + \Delta^{\mu \nu} \partial_{\alpha}\pi^{\alpha}_{ \ \nu} &= 0,
\end{align}
\end{subequations}
where we make use of the irreducible decomposition of the derivative, $\partial_{\mu} = u_{\mu}D + \nabla_{\mu}$, which defines the co-moving derivative $D \equiv u^{\nu} \partial_{\nu}$, and the space-like gradient $\nabla_{\mu} = \Delta_{\mu}^{\nu} \partial_{\nu}$. 

\subsection*{Collision term approximation}

In this work, we express the collision term of the effective kinetic theory using the new relaxation time approximation (RTA) described in Ref.~\cite{rocha:2021}. The collision term is first approximated as a linear collision operator acting on the non-equilibrium correction $\phi_\pp$, as $C[f_\pp] \approx f_{0\pp}\hat{L}\phi_\pp$. The linear operator $\hat{L}$ is then approximated to have the following general structure,
\begin{equation}
    \hat{L}  \propto  \mathds{1} -  \sum_{n} \vert \lambda_{n} \rangle \langle \lambda_{n} \vert,
    \label{defineC}
\end{equation}
where $\lambda_{n}$'s are the degenerate orthogonal eigenvectors with zero eigenvalue associated with the microscopically conserved quantities. In contrast to the traditional Anderson-Witting approximation \cite{ANDERSON1974466} (where $\hat{L}  \propto  \mathds{1}$), the Ansatz in \eqref{defineC} guarantees that macroscopic laws for energy-momentum conservation emerge regardless of the choice of matching condition (hydrodynamic frame) or energy dependence of the relaxation time.    
In the present case, we intend to model a gas of vanishing net charge and the only relevant associated macroscopic conserved quantities are energy and momentum. Hence, we do not need to consider the counter-terms in \eqref{defineC} associated to particle number conservation, as was originally considered in Ref.~\cite{rocha:2021}. This leads to the following approximation to the collision operator,
\begin{equation}
\label{eq:nRTA}
\begin{aligned}
f_{0\textbf{p}}\hat{L}\phi_{\mathbf{p}} \equiv - \frac{E_{\mathbf{p}}}{\tau_{R}} f_{0\textbf{p}} \left[ \phi_{\mathbf{p}} - 
\frac{\lan \phi_{\mathbf{p}}  \frac{E_{\mathbf{p}}^{2}}{\tau_{R}} \ran_{0}}{\lan  \frac{E_{\mathbf{p}}^{3}}{\tau_{R}}\ran_{0}} E_{\bf p} 
- 
\frac{\lan \phi_{\mathbf{p}}  \frac{E_{\mathbf{p}}}{\tau_{R}} p^{\langle \mu \rangle} \ran_{0}}{ \frac{1}{3}\lan \delpp  \frac{E_{\mathbf{p}}}{\tau_{R}}\ran_{0}} p_{\langle \mu \rangle}   \right],
\end{aligned}
\end{equation}
where we introduce the relaxation time, $\tau_R$. The energy dependence of the relaxation time is parametrized as, 
\begin{equation}
\label{eq:tR-param}
    \tau_{R} = t_{R} \left( \frac{E_{\mathbf{p}}}{T} \right)^{\gamma},
\end{equation}
where $t_R$ has no energy dependence. This is more general than the usual assumption of a constant (energy-independent) relaxation time so that by varying $\gamma$ one can investigate other functional forms that are more phenomenologically relevant for the problem at hand. Indeed, it is commonly argued that in QCD effective kinetic theories  $\gamma = 1/2$ \cite{dusling2010radiative}, while in scalar field theories $\gamma = 1$ \cite{calzetta-88:nonequilibrium}.

\section{Chapman-Enskog expansion}
\label{sec:coeffs}

In this section, we use approximation \eqref{eq:nRTA} to calculate the fluid-dynamical transport coefficients following the traditional Chapman-Enskog expansion \cite{chapman1916vi,chapman1990mathematical,enskog1917kinetische,deGroot:80relativistic}. In this case, a perturbative solution of the Boltzmann equation is obtained in which the single-particle distribution function is expanded in a series of powers of space-like gradients. In practice, this solution is constructed by inserting a book-keeping parameter $\epsilon$ in all terms of the Boltzmann equation that contain a derivative, 
\begin{equation}
\label{eq:BoltzmannQuasiPert}
\epsilon \left( p^{\mu} \partial_\mu f_\pp + \frac{1}{2}\partial_i M^2(T) \partial^i_{(\pp)} f_\pp \right) = f_{0\textbf{p}}\hat{L}\phi_{\mathbf{p}} ,
\end{equation}
and expanding the single-particle distribution function in powers of $\epsilon$, 
\begin{equation}
\begin{aligned}
\label{eq:expn-knudsen}
f_{\mathbf{p}} = \sum_{i=0}^{\infty} \epsilon^{i} f^{(i)}_{\mathbf{p}}.
\end{aligned}
\end{equation}
This expansion is used in Eq.~\eqref{eq:BoltzmannQuasiPert} and solved order-by-order in $\epsilon$. In the end, an asymptotic solution of the Boltzmann equation is recovered by truncating the series and setting $\epsilon=1$.

The zeroth order solution resulting from this procedure is the local equilibrium distribution function, $f^{(0)}_\pp = f_{0 \pp}$. The first order solution is obtained by solving the following  equation for $\phi_\pp^{ (1)} = f_{\pp}^{(1)}/ f_{0\pp}$ \cite{deGroot:80relativistic},
\begin{equation}
\begin{aligned}
\label{eq:integral-nRTA}
 \left( A_{\mathbf{p}} \theta 
- 
\beta p^{\langle \mu}p^{\nu \rangle} \sigma_{\mu \nu} \right)f_{0\mathbf{p}}
= 
 - \frac{E_{\mathbf{p}}}{\tau_{R}} f_{0\textbf{p}} \left[ \phi_{\mathbf{p}}^{(1)} - 
\frac{\lan \phi_{\mathbf{p}}^{(1)}  \frac{E_{\mathbf{p}}^{2}}{\tau_{R}} \ran_{0}}{\lan  \frac{E_{\mathbf{p}}^{3}}{\tau_{R}}\ran_{0}}   E_{\bf p}
- 
\frac{\lan \phi_{\mathbf{p}}^{(1)}  \frac{E_{\mathbf{p}}}{\tau_{R}} p^{\langle \mu \rangle} \ran_{0}}{ \frac{1}{3}\lan \delpp  \frac{E_{\mathbf{p}}}{\tau_{R}}\ran_{0}} p_{\langle \mu \rangle}   \right]
,
\end{aligned}
\end{equation}
where we defined
\begin{subequations}
\begin{align}
\label{eq:Ap}
& A_{\mathbf{p}} = -  \beta c_{s}^{2} E_{\bf p}^{2} - \frac{\beta}{3} \Delta^{\lambda \sigma}p_{\lambda} p_{\sigma} 
-
\beta^{2} M \frac{\partial M}{\partial \beta} c_{s}^{2} ,
\\
\label{eq:cs^2}
& c_{s}^{2} \equiv \left(\frac{\partial P_{0}}{\partial \varepsilon_{0}} \right) =  
\frac{I_{31}}{I_{30} + \frac{1}{2} I_{10}\frac{\partial M^{2}}{\partial \beta} \beta},
\end{align}    
\end{subequations}
and made use of the following notation
\begin{equation}
\label{eq:irrep-coeffs}
\begin{aligned}
&  I_{nq} = \frac{1}{(2q+1)!!} \left\langle \left( -\Delta^{\lambda \sigma}p_{\lambda} p_{\sigma} \right)^{q} E_{\mathbf{p}}^{n-2q}\right\rangle_{0}. 
\end{aligned}    
\end{equation}
We note that, when deriving Eq.\ \eqref{eq:integral-nRTA}, the time-like derivatives of $f_{0\pp}$ were replaced by space-like ones using
\begin{equation}
\label{eq:df/dt}
\begin{aligned}
D f_{0\mathbf{p}} 
=
\left[ - \beta c_{s}^{2} E_{\bf p} \theta   -  \frac{\beta}{\varepsilon_{0} + P_{0}} p_{\langle \mu \rangle} \nabla^{\mu}P_{0}\right] f_{0\bf p},
\end{aligned}    
\end{equation}
together with the Gibbs-Duhem relation \cite{landau2013statistical} applied at zero chemical potential
\begin{equation}
\begin{aligned}
\label{eq:Gibbs-duhem}
\nabla_{\mu} \beta = - \frac{\beta }{\varepsilon_{0}+ P_{0}} \nabla_{\mu} P_{0}.
\end{aligned}
\end{equation}

The solution of Eq.~\eqref{eq:integral-nRTA} is of the form
\begin{equation}
\label{eq:phi1-hom-part}
\phi_{\pp}^{(1)} = \phi_{\pp}^{\rm{hom}} + \phi_{\pp}^{\rm{part}},      
\end{equation}
where $\phi_{\pp}^{\rm{hom}}$ denotes the homogeneous solution ($\hat{L}\phi_{\mathbf{p}}^{\rm{hom}} = 0$), while  $\phi_{\pp}^{\rm{part}}$ is the particular solution. The homogeneous solution is a linear combination of the collisional invariants, hence
\begin{equation}
\phi_{\pp}^{\rm{hom}} = a_{\mu} p^{\mu},   
\end{equation}
where $a_{\mu}$ is determined using the matching conditions \eqref{eq:match1} and \eqref{eq:matching_kinetic1} \cite{deGroot:80relativistic}. In order to determine the particular solution, one expands   $\phi_{\mathbf{p}}^{\rm{part}}$ in an irreducible orthogonal basis \cite{deGroot:80relativistic}
\begin{equation}
\label{eq:expn-tensorial}
\begin{aligned}
\phi^{\rm{part}}_{\mathbf{p}} = \sum_{\ell,n = 0}^{\infty}  \Phi_{n}^{\langle \mu_{1} \cdots \mu_{\ell} \rangle}  p_{\langle \mu_{1}} \cdots  p_{\mu_{\ell} \rangle}P_{n}^{(\ell)}(E_\pp),
\end{aligned}
\end{equation}
where $p^{\langle \mu_{1}} \cdots  p^{\mu_{\ell} \rangle}$, $\ell=0,1,\ldots$, are irreducible tensors constructed from $p^{\mu}$ \cite{deGroot:80relativistic}, and $P_{n}^{(\ell)}$ are orthogonal polynomials of $\beta E_\pp$, with $n=0,1,\ldots$. For the purposes of this paper, it is sufficient to know that the irreducible tensors satisfy the following orthogonality condition,
\begin{equation}
\label{eq:tensor-rel}
\begin{aligned}
\int dP\,
p^{\langle \mu_{1}} \cdots p^{\mu_{\ell} \rangle}
p_{\langle \nu_{1}} \cdots p_{\nu_{m} \rangle}
H(E_{\mathbf{p}})
& = \frac{\ell!\delta_{\ell m}}{(2\ell + 1)!!}  \Delta^{\mu_{1} \cdots \mu_{\ell}}_{\nu_{1} \cdots \nu_{\ell}} \int dP  \left(\Delta^{\mu \nu} p_{\mu} p_{\nu} \right)^{\ell} H(E_{\mathbf{p}}),  
\end{aligned}
\end{equation}
with $H(E_{\mathbf{p}})$ being an arbitrary function of $E_\pp$   (assumed to fall sufficiently fast with $E_\pp$ to ensure the convergence of the integrals). On the other hand, the polynomials $P_{n}^{(\ell)}$ are constructed to obey the orthogonality relations, 
\begin{equation}
\begin{aligned}
\label{eq:polys-orthogonal-2}
\left( P^{(\ell)}_{m} ,  P^{(\ell)}_{n}   \right)_{\ell} 
&= A^{(\ell)}_{m} \delta_{mn}, 
\end{aligned}
\end{equation}
with $A_{m}^{(\ell)}$ being a normalization factor. We define the inner product used in \eqref{eq:polys-orthogonal-2} as follows
\begin{equation}
\label{eq:inner-prod-l}
\begin{aligned}
\left( \phi , \psi \right)_{\ell} 
= 
\frac{\ell!}{(2\ell + 1)!!} \left\langle \left( \Delta^{\mu \nu} p_{\mu} p_{\nu} \right)^{\ell} \frac{E_{\mathbf{p}}}{\tau_{R}}   \phi_{\mathbf{p}} \psi_{\mathbf{p}} \right\rangle_{0}.
\end{aligned}    
\end{equation}
For the case of massless particles, the basis of orthogonal polynomials introduced above can be identified as associated Laguerre polynomials. For massive particles, these polynomials do not correspond to any known special function and they have to be constructed using, e.g. the Gram-Schmidt algorithm. In this case, we impose that $P^{(0)}_{0} = \beta E_{\bf p}$, $P^{(\ell \geq 1)}_{0} = 1$, and determine
the subsequent elements of the polynomial basis using the orthogonality condition \eqref{eq:polys-orthogonal-2}. 

Substituting expansion \eqref{eq:expn-tensorial} in Eq.~\eqref{eq:integral-nRTA}, and using the
orthogonality relations introduced above, we obtain the following solution for the expansion coefficients $\Phi_n$
\begin{subequations}
\label{eq:soln-Phis}
\begin{align}
\label{eq:soln-Phis-l=0}
& \Phi_{n} =  - \frac{\theta}{A_{n}^{(0)}} \left\langle A_{\mathbf{p}} P^{(0)}_{n} \right\rangle_{0} \; (n \geq 1), \\
\label{eq:soln-Phis-l=1}
& \Phi_{n}^{\mu} =  0 \; (n \geq 1), \\
& \label{eq:soln-Phis-l=2}
\Phi_{n}^{\mu \nu} = \beta \frac{\sigma^{\mu \nu}}{A_{n}^{(2)}} \left\langle \left(\Delta^{\mu \nu} p_{\mu} p_{\nu}\right)^{2}  P^{(2)}_{n} \right\rangle_{0} \; (n \geq 0)\\
&
\label{eq:soln-Phis-l>2}
\Phi_{n}^{\mu_{1} \cdots \mu_{\ell}} = 0
\ (\ell \geq 3, n \geq 0). 
\end{align}
\end{subequations}
The coefficients $\Phi_{0}$ and $\Phi_{0}^{\mu}$ are removed from this inversion procedure since they are incorporated into the homogeneous solution of $\phi^{(1)}_{\pp}$. Without loss of generality, this procedure can be carried out by redefining the coefficients of the homogeneous solution,  $a_{\mu}$, as
\begin{equation}
\begin{aligned}
& \Phi_{0} = \frac{a_{\mu}u^{\mu}}{\beta}, \\
& \Phi_{0}^{\mu} = a^{\lan \mu \ran}.
\end{aligned}    
\end{equation}
The components related to the homogeneous solution are then found by enforcing the matching conditions \eqref{eq:match1} and \eqref{eq:matching_kinetic1}, which yields
\begin{equation}
\label{eq:matching-phis}
\begin{aligned}
& 
\Phi_{0}
=
- \sum_{n=1}^{\infty} \Phi_{n} \frac{\langle P_{n}^{(0)} \rangle_{0}}{\langle P_{0}^{(0)} \rangle_{0}}, \\
&
\Phi_{0}^{\mu} = 0.
\end{aligned}    
\end{equation}
Finally, collecting results \eqref{eq:soln-Phis} and \eqref{eq:matching-phis} in Eq.~\eqref{eq:phi1-hom-part} and summing the series (see Appendix \ref{sec:sum-coeffs}), we find 
\begin{equation}
\label{eq:solution-phi}
\begin{aligned}
& \phi^{(1)}_{\mathbf{p}}= F^{(0)}_{\mathbf{p}} \theta
+
F^{(2)}_{\mathbf{p}} p^{\langle \mu}  p^{\nu \rangle} \sigma_{\mu \nu},
\end{aligned}    
\end{equation}
where we defined the scalar functions of $E_\pp$
\begin{subequations}
\begin{align}
&
\label{eq:F0-sol}
F^{(0)}_{\mathbf{p}} = 
-
\frac{\tau_{R}}{E_{\mathbf{p}}} A_{\pp}
+
\frac{1}{I_{1,0}}
\left\langle \frac{\tau_{R}}{E_{\mathbf{p}}} A_{\pp}  \right\rangle_{0}  E_{\bf p},
\\
&
\label{eq:F2-sol}
F^{(2)}_{\mathbf{p}} =  \beta \frac{\tau_{R}}{E_{\mathbf{p}}},   
\end{align}    
\end{subequations}
and used the fact that $\langle A_{\bf p} E_{\bf p} \rangle = 0$. We note that there is no vector component for $\phi_{\bf p}^{(1)}$, as a result of the vanishing chemical potential assumption. Also, we remark  that if a  matching condition different than \eqref{eq:match1} were used, only the second term of $F^{(0)}_\pp$ would be modified.

\begin{figure}[!h]
    \centering
\begin{subfigure}{0.47\textwidth}
\label{fig:F0a50MeV}
  \includegraphics[width=\linewidth]{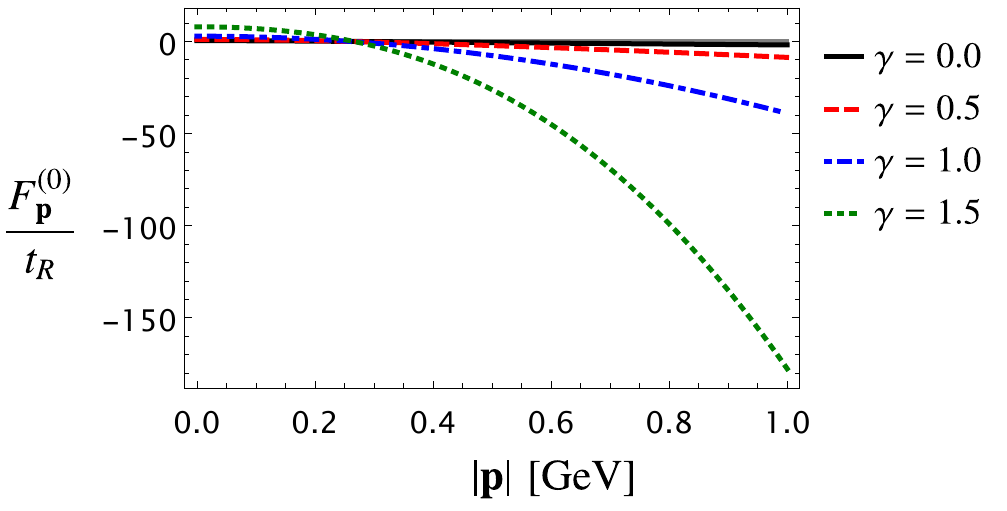}
\end{subfigure}\hfil 
\begin{subfigure}{0.47\textwidth}
\label{fig:F2b50MeV}
  \includegraphics[width=\linewidth]{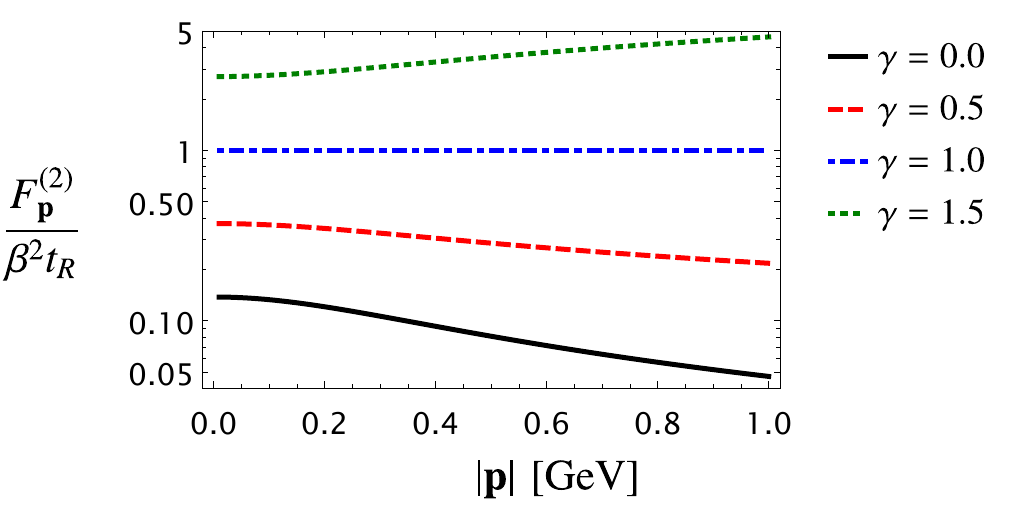}
\end{subfigure}\hfil 

\caption{The scalar (left panel) and tensor (right panel) components of the non-equilibrium corrections to the single particle distribution function as a function of the momentum of the particle in the local rest frame of the fluid. Results are shown for $T = 50$ MeV, for several values of the phenomenological parameter $\gamma$ [cf.~Eq.~\eqref{eq:tR-param}].}
\label{fig:F0-F2-50MeV}
\end{figure}

\begin{figure}[!h]
    \centering
\begin{subfigure}{0.47\textwidth}
\label{fig:F0a}
  \includegraphics[width=\linewidth]{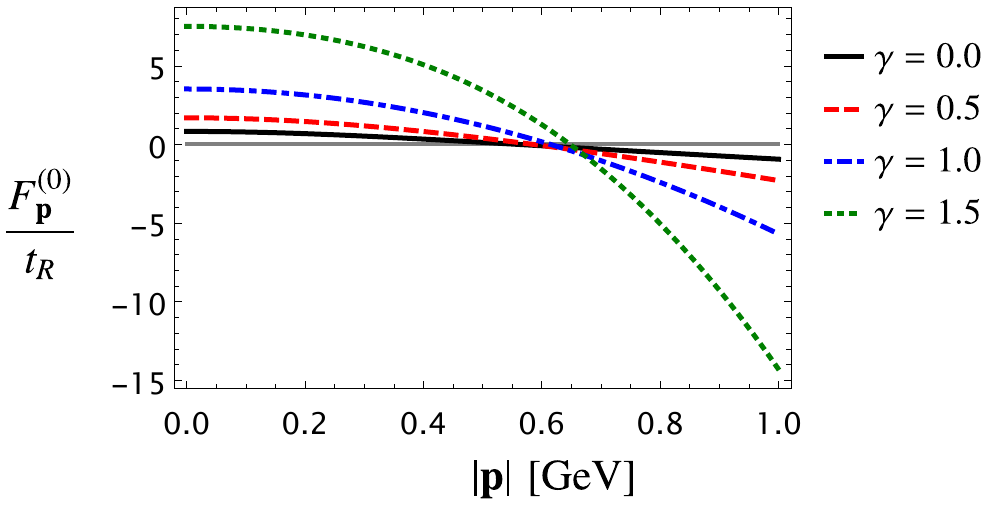}
\end{subfigure}\hfil 
\begin{subfigure}{0.47\textwidth}
\label{fig:F2b}
  \includegraphics[width=\linewidth]{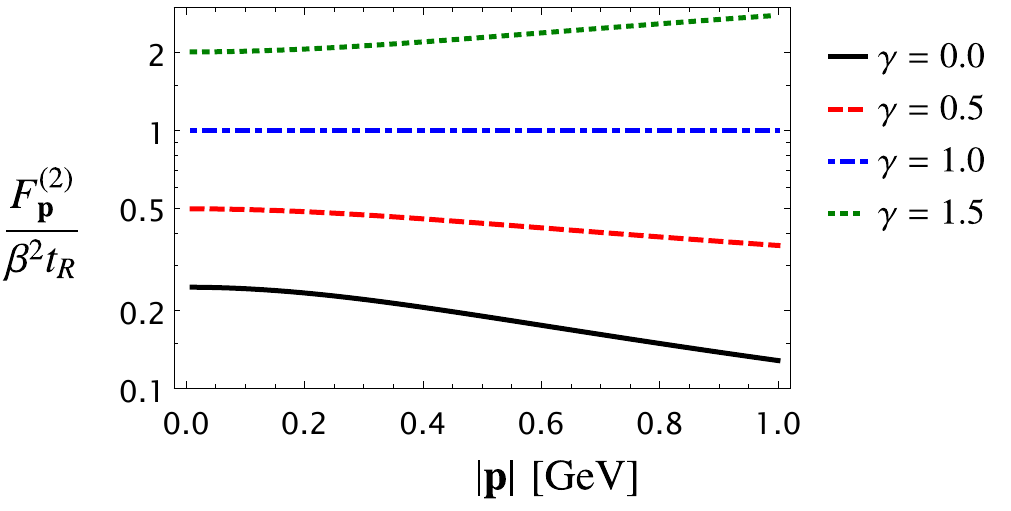}
\end{subfigure}\hfil 

\caption{The scalar (left panel) and tensor (right panel) components of the non-equilibrium corrections to the single particle distribution function as a function of the momentum of the particle in the local rest frame of the fluid. Results are shown for $T = 150$ MeV, for several values of the phenomenological parameter $\gamma$ [cf.~Eq.~\eqref{eq:tR-param}].}
\label{fig:F0-F2}
\end{figure}

\begin{figure}[!h]
    \centering
\begin{subfigure}{0.47\textwidth}
\label{fig:F0a250MeV}
  \includegraphics[width=\linewidth]{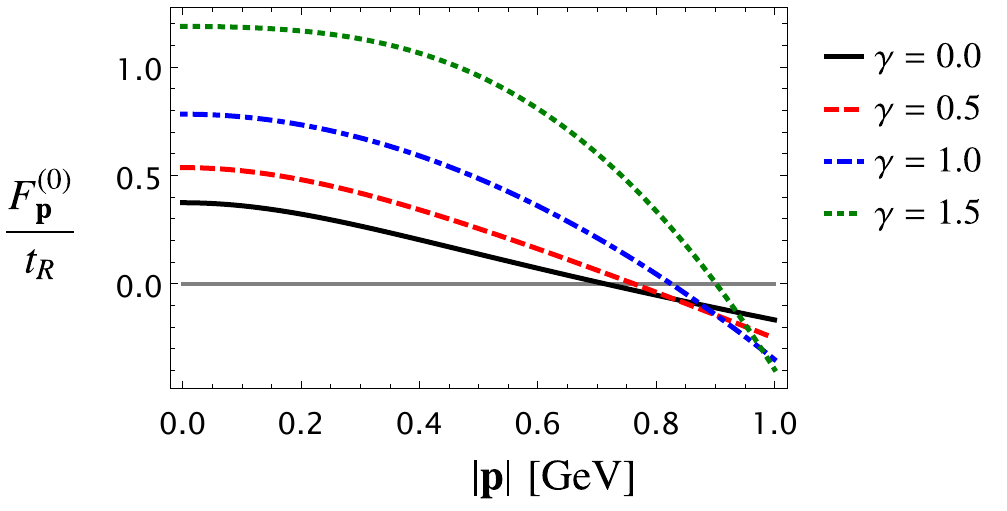}
\end{subfigure}\hfil 
\begin{subfigure}{0.47\textwidth}
\label{fig:F2b250MeV}
  \includegraphics[width=\linewidth]{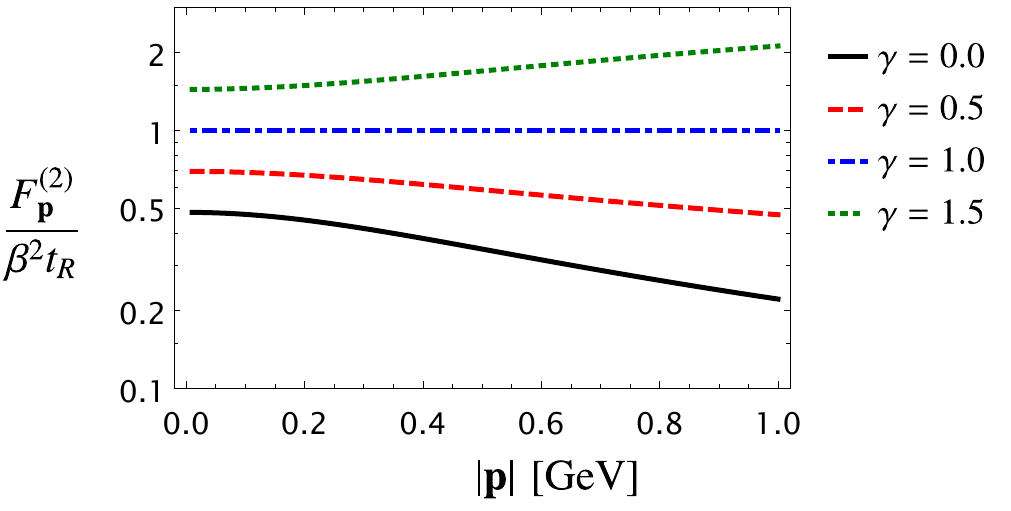}
\end{subfigure}\hfil 

\caption{The scalar (left panel) and tensor (right panel) components of the non-equilibrium corrections to the single particle distribution function as a function of the momentum of the particle in the local rest frame of the fluid. Results are shown for $T = 250$ MeV, for several values of the phenomenological parameter $\gamma$ [cf.~Eq.~\eqref{eq:tR-param}].}
\label{fig:F0-F2-250MeV}
\end{figure}

In Figs.~\ref{fig:F0-F2-50MeV}, \ref{fig:F0-F2} and \ref{fig:F0-F2-250MeV}, we plot $F_\pp^{(0,2)}$ as functions of the particle momentum in the local rest frame of the fluid, for three different temperatures, $T = 50$ MeV, $150$ MeV and $250$ MeV. These plots also show how $F_\pp^{(0)}$ and $F_\pp^{(2)}$ vary with the phenomenological parameter $\gamma$, which determines the temperature dependence of the relaxation time. One can see that both functions are significantly affected by the parameter $\gamma$. In particular the scalar component, $F^{(0)}_\pp$. We note that the tensor component, $F^{(2)}_\pp$, becomes constant for $\gamma = 1$, a behavior expected for scalar field theories \cite{calzetta-88:nonequilibrium}.

\section*{Transport coefficients}


Using definitions \eqref{eq:def_kinetic} and \eqref{eq:def_kinetic2} of the non-equilibrium currents and the first order solution of the single particle distribution function derived above, we obtain the following constitutive relations
\begin{equation}
\label{eq:def-coeffs}
\begin{aligned}
\Pi & \simeq - \frac{1}{3} \left\langle \left(\Delta^{\mu \nu} p_{\mu} p_{\nu}\right) \phi^{(1)}_{\mathbf{p}} \right\rangle_{0}  \equiv - \zeta \theta,  \\
 \delta \varepsilon & \simeq \left\langle E_{\mathbf{p}}^{2}  \phi^{(1)}_{\mathbf{p}} \right\rangle_{0} \equiv \chi \theta, \\ 
\pi^{\mu \nu} & \simeq \left\langle p^{\langle \mu} p^{ \nu \rangle} \phi^{(1)}_{\mathbf{p}} \right\rangle_{0} \equiv  2 \eta \sigma^{\mu \nu},
\end{aligned}
\end{equation}
with transport coefficients given by
\begin{equation}
\label{eq:coeffs-all}
\begin{aligned}
& \zeta =  -\frac{1}{3}  \left\langle \left(\Delta^{\mu \nu} p_{\mu} p_{\nu}\right) A_{\mathbf{p}} \frac{\tau_{R}}{E_{\mathbf{p}}} \right\rangle_{0} 
-
\left\langle \frac{\tau_{R}}{E_{\mathbf{p}}} A_{\mathbf{p}} \right\rangle_{0} \frac{I_{3,1}}{I_{1,0}},  
\\
& \chi = - \left\langle  A_{\mathbf{p}} \tau_{R} E_{\mathbf{p}} \right\rangle_{0} 
+ 
\left\langle \frac{\tau_{R}}{E_{\mathbf{p}}} A_{\mathbf{p}} \right\rangle_{0} \frac{I_{3,0}}{I_{1,0}}  , \\ 
& \eta = \frac{\beta}{15} \left\langle \left(\Delta^{\mu \nu} p_{\mu} p_{\nu}\right)^{2}\frac{\tau_{R}}{E_{\mathbf{p}}} \right\rangle_{0} .
\end{aligned}
\end{equation}
%
%

For the sake of completeness, below we obtain simplified expressions for these transport coefficients in two distinct regimes. First, in the limit $M(T)/T \rightarrow 0$ the asymptotic behavior for these transport coefficients is
\begin{equation}
\label{eq:coeffs-m=0}
\begin{aligned}
&
\frac{\zeta}{t_{R}(\varepsilon_{0}+ P_{0})} \sim - \frac{1}{864} M(T)  \frac{d}{dT}\left(\frac{M(T)}{T}\right)(\gamma ^4+10 \gamma ^3+11 \gamma ^2-22 \gamma +120)\Gamma(\gamma+1) ,\\
&
\frac{\chi}{t_{R}(\varepsilon_{0}+ P_{0})} \sim 
\frac{1}{288} M(T) \frac{d}{dT}\left(\frac{M(T)}{T}\right) \left( \gamma ^4+10 \gamma ^3+11 \gamma ^2-22 \gamma +120 \right)\Gamma(\gamma+1),\\
&
\frac{\eta}{t_{R}(\varepsilon_{0}+ P_{0})} \sim \frac{\Gamma(\gamma + 5)}{120},
\end{aligned}    
\end{equation}
where $\Gamma(z)$ denotes the Gamma function \cite{gradshteyn2014table}. The results above are consistent with the fact that $\zeta$ and $\chi$ must vanish when the particle mass is zero. Indeed, from equation \eqref{eq:cs^2}, we obtain that, in this limit  
\begin{equation}
\label{eq:cs2-exp-m=0}
\begin{aligned}
& c_{s}^{2} \sim \frac{1}{3} + \frac{M(T)}{36} \frac{d}{dT}\left(\frac{M(T)}{T}\right).
\end{aligned}    
\end{equation}
Hence, at leading order in $M(T)/T$, $\zeta,\chi \propto (c_{s}^{2} - 1/3)$, which differs from the behavior $\zeta \propto (c_{s}^{2} - 1/3)^{2}$, expected in this limit. In the next section, we show that this property is only attained by the matching-invariant coefficient $\zeta_{s}$, defined from entropy production. We also note that the shear coefficient reduces to the well known RTA result $\eta = (1/5) t_{R}(\varepsilon_{0}+ P_{0})$, for $\gamma = 0$. Moreover, one can see that all coefficients grow factorially with $\gamma$. On the other hand, when $M(T)/T \rightarrow \infty$, the asymptotic behavior is
\begin{equation}
\label{eq:coeffs-m-inf}
\begin{aligned}
&
\frac{\zeta}{t_{R}(\varepsilon_{0} +P_{0})}
\sim 
- \frac{2}{3} \left(\frac{M(T)}{T}\right)^{\gamma - 1} (3 c_{s}^{2} -1),
\\
&
\frac{\chi}{t_{R}(\varepsilon_{0} +P_{0})} \sim 
\left(\frac{M(T)}{T}\right)^{\gamma-1}2(3c_{s}^{2}-1),
\\
&
\frac{\eta}{t_{R}(\varepsilon_{0}+ P_{0})} \sim  \left(\frac{M(T)}{T}\right)^{\gamma - 1}, 
\end{aligned}    
\end{equation}
and, hence, all transport coefficients behave asymptotically with the same power of the normalized thermal mass. Furthermore, we note that for the thermal mass used in this paper, a regime in which $M(T)/T \ll 1 $ is never really reached. The smallest values of mass are obtained in the large temperature regime but, even then, with $M(T)/T$ always being larger than one. Indeed, it can be shown from Eqs.~\eqref{eq:s_K}, \eqref{eq:WP_tr_anom} and \eqref{eq:P_from_I} that $M(T)/T \to 1.1$ as $T \to \infty$. On the other hand, large $M(T)/T$ values are reached as the temperature approaches zero. It is also noted from Eqs.~\eqref{eq:coeffs-m=0} and \eqref{eq:coeffs-m-inf} that in both regimes $\chi = - 3 \zeta$.

\begin{figure}[!h]
    \centering
\begin{subfigure}{0.48\linewidth}
  \includegraphics[width=\linewidth]{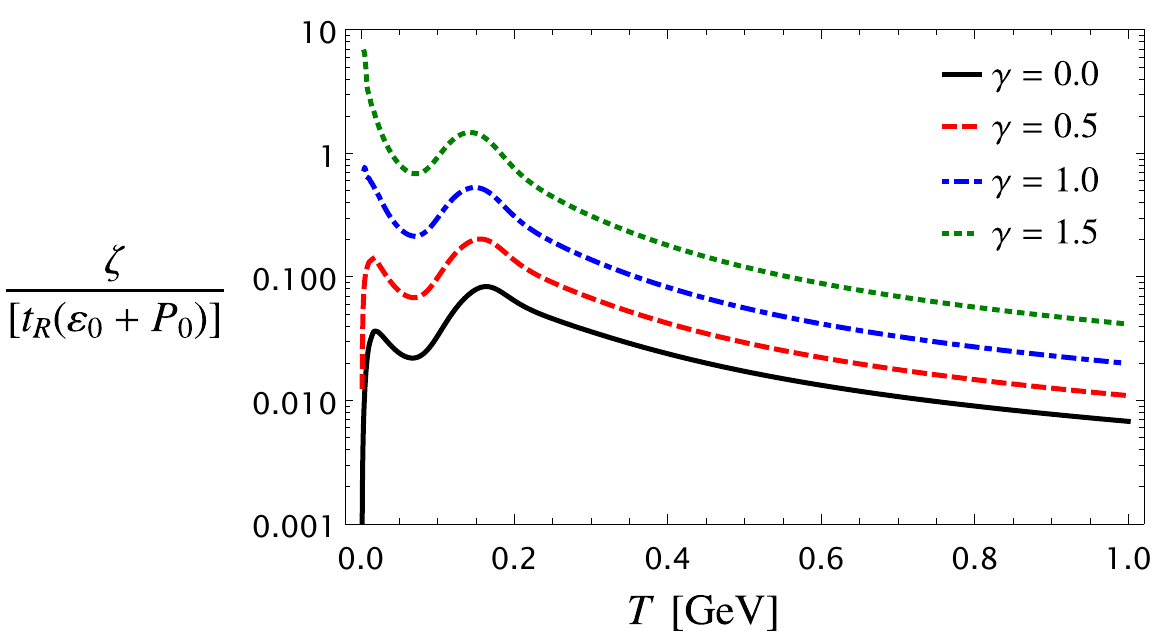}
  \label{fig:zeta-gammas}
\end{subfigure}\hfil 
\begin{subfigure}{0.48\linewidth}
  \includegraphics[width=\linewidth]{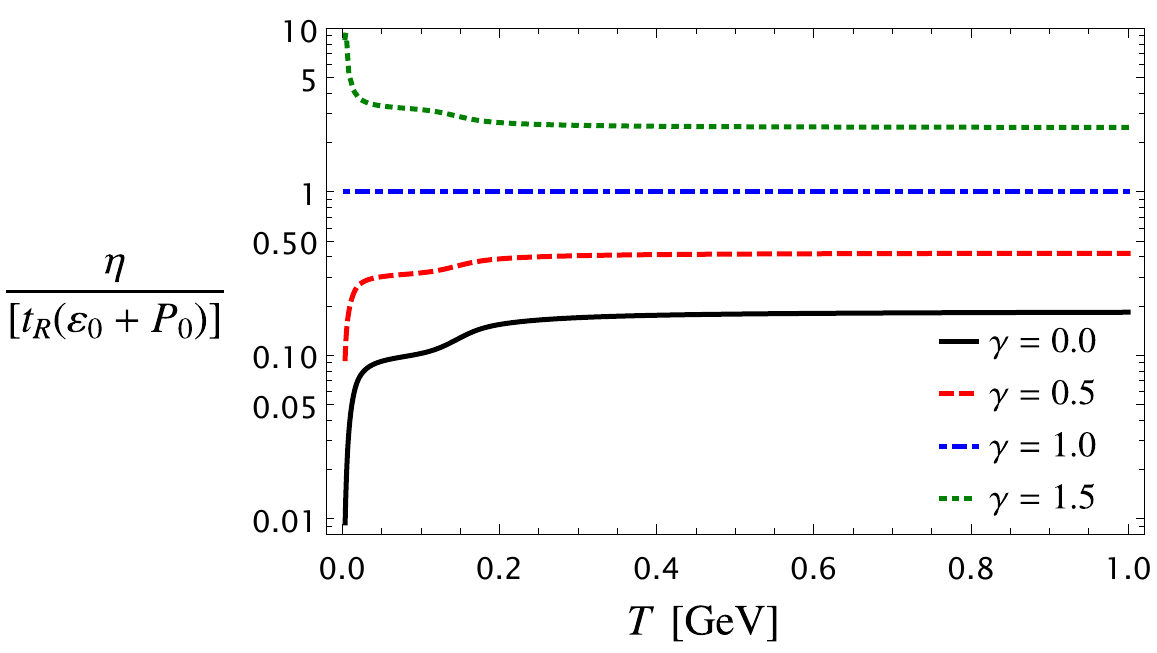}
  \label{fig:eta-gammas}
\end{subfigure}\hfil
\\
\begin{subfigure}{0.48\linewidth}
  \includegraphics[width=\linewidth]{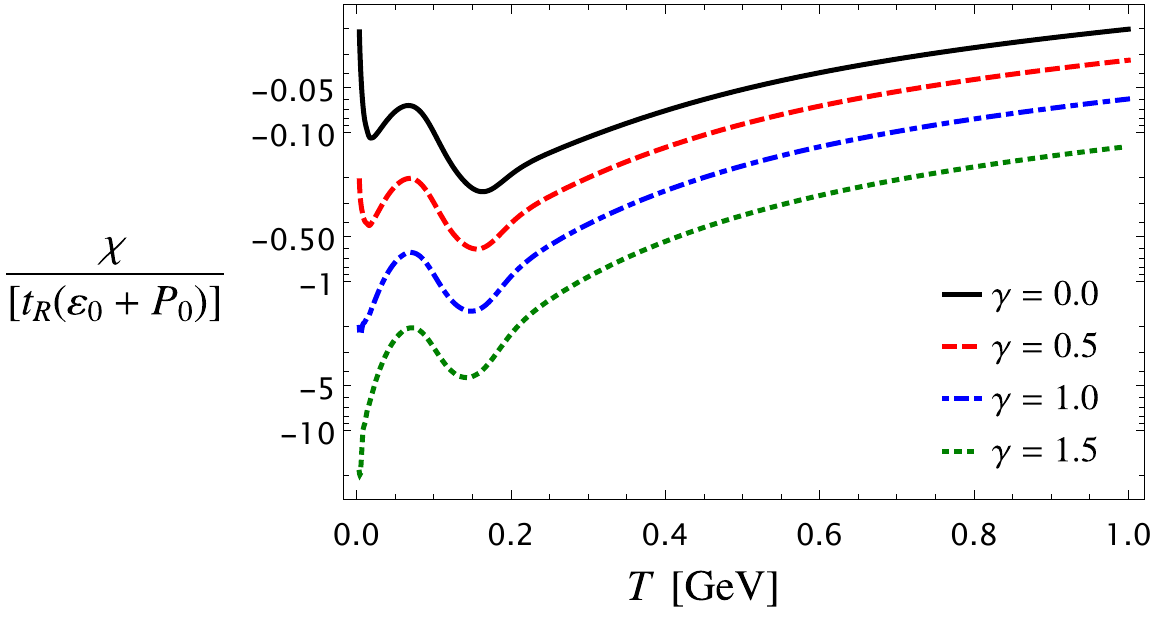}
  \label{fig:chi-gammas}
\end{subfigure}\hfil 
\caption{Normalized transport coefficients for the quasiparticle model as a function of temperature for various values of the phenomenological parameter $\gamma$. (Top left) Normalized bulk viscosity. (Top right) Normalized shear viscosity. (Bottom) Normalized energy deviation coefficient.}
\label{fig:eta-zeta-gammas}
\end{figure}

In Fig.~\ref{fig:eta-zeta-gammas} we plot all transport coefficients as functions of temperature for different phenomenological parameters $\gamma$ (cf.~\eqref{eq:tR-param}). We find that the effect of the energy dependence of the relaxation time on the transport coefficients is significant. In particular, we see that the overall magnitude of all transport coefficients increases with $\gamma$. We further note that the curves for the energy deviation coefficient, $\chi$, which parametrize the out of equilibrium corrections to the energy density, are similar to those for the bulk viscosity, $\zeta$, but with an opposite sign and being larger in modulus.    

\section{Entropy production}
\label{sec:ent-prod}

In this section we study the entropy production of the effective kinetic theory discussed in this paper in the fluid-dynamical regime. In kinetic theory, one usually defines the entropy current for classical quasiparticles as \cite{deGroot:80relativistic}
\begin{equation}
\begin{aligned}
S^{\mu} = \int dP\, p^{\mu} f_{\bf p}( 1 - \ln f_{\bf p} ).
\end{aligned}
\end{equation}
We first demonstrate that this expression has a non-negative four-divergence, even with the addition of the mass-dependent, Vlasov-like term in Eq.~\eqref{eq:BoltzmannQuasi}. In fact, using the Boltzmann equation \eqref{eq:BoltzmannQuasi}, we have that
\begin{equation}
\begin{aligned}
\partial_{\mu} S^{\mu} = - \int dP \left[ \left( C[f] - \frac{1}{2} \partial_{i}M^{2} \partial^{i}_{(\bf{p})}f_{\bf p} \right) \ln f_{\bf p} 
+
\frac{1}{2 E_{\bf p}^{2}} \partial_{\mu}M^{2} p^{\mu} f_{\bf p}( 1 - \ln f_{\bf p} )
-
\partial_{\mu}p^{\mu} f_{\bf p}( 1 - \ln f_{\bf p} )\right].
\end{aligned}
\end{equation}
The last two terms partially cancel each other due to the fact that $\partial_{\mu}p^{\mu} = [1/(2 E_{\bf p})] \partial_{0}M^{2}$. The remaining term cancels with the second term inside the brackets because 
\begin{equation}
\begin{aligned}
&
\int dP  \frac{1}{2} \partial_{i}M^{2} \partial^{i}_{(p)}f_{\bf p} \ln f_{\bf p} 
= 
\int dP  \frac{1}{2} \partial_{i}M^{2} \partial^{i}_{(p)}(f_{\bf p} \ln f_{\bf p} - f_{\bf p}) 
=
- 
\int dP \frac{p^{i}}{E_{\bf p}^{2}}  \frac{1}{2} \partial_{i}M^{2} (f_{\bf p} \ln f_{\bf p} - f_{\bf p}), 
\end{aligned}    
\end{equation}
where, in the last part, an integral by parts was performed. Then, we recover the usual identity
\begin{equation}
\begin{aligned}
\partial_{\mu} S^{\mu} = - \int dP\, C[f_{\mathbf{p}}] \ln f_{\mathbf{p}}.
\end{aligned}
\end{equation}

In leading order in the book-keeping parameter $\epsilon$ introduced in the last section, the entropy production can be simplified to  
\begin{equation}
\partial_{\mu} S^{\mu}  \simeq - \epsilon \int dP \phi^{(1)}_{\mathbf{p}} \hat{L}[\phi^{(1)}_{\mathbf{p}}] f_{0\mathbf{p}}, 
\end{equation}
which, imposing the relaxation time approximation \eqref{eq:nRTA} and using the moment expansion for $\phi^{(1)}$ \eqref{eq:expn-tensorial}, becomes 
\begin{equation}
\begin{aligned}
\partial_{\mu} S^{\mu} \simeq   \sum_{n=1}^{\infty} A^{(0)}_{n} \Phi_{0n}^{2} 
+
\frac{1}{3}\sum_{n=1}^{\infty} A^{(1)}_{n} \Phi_{n}^{\mu}\Phi_{n \mu}  
+
\frac{2}{15}\sum_{n=0}^{\infty} A^{(2)}_{n} \Phi_{n}^{ \mu \nu}\Phi_{n \mu \nu}  
+
\sum_{\ell = 3}^{\infty}\frac{\ell!}{(2\ell + 1)!!}\sum_{n=0}^{\infty} A^{(\ell)}_{n} \Phi_{n}^{ \mu_{1} \cdots \mu_{\ell} }\Phi_{n  \mu_{1} \cdots \mu_{\ell} }.
\end{aligned}    
\end{equation}
This expression is manifestly non-negative due to the fact that $ (-1)^{\ell} \Phi_{n}^{ \mu_{1} \cdots \mu_{\ell} }\Phi_{n  \mu_{1} \cdots \mu_{\ell} } \geq 0$ and $ (-1)^{\ell} A^{(\ell)}_{n} \geq 0$ (cf.~Eq.~\eqref{eq:polys-orthogonal-2}) and, thus, is consistent with the second law of thermodynamics. Furthermore, we note that the entropy production does not depend on the expansion coefficients $\Phi_{0}$ and $\Phi_{0}^{\mu}$, which contain all the dependence on the matching conditions imposed. This implies that the entropy production, at least for the linearized collision kernel, is unaffected by the choice of matching condition.  

Substituting the first order solution for the coefficients $\Phi_n$ obtained in the last section, the entropy production can be expressed in the following form,
\begin{equation}
\label{eq:zeta-s}
\begin{aligned}
& 
\partial_{\mu} S^{\mu} 
\simeq 
\zeta_{s} \theta^{2}
+
2 \eta_{s} \sigma^{\mu \nu} \sigma_{\mu \nu},\\
&
\zeta_{s} =  \left\langle \frac{\tau_{R}}{E_{\bf p}} [A_{\bf p}]^{2} \right\rangle_{0}, \\
&
\eta_{s} =  \frac{\beta}{15} \left\langle \left(\Delta^{\mu \nu} p_{\mu} p_{\nu}\right)^{2}\frac{\tau_{R}}{E_{\mathbf{p}}} \right\rangle_{0}.
\end{aligned}    
\end{equation}
It is straightforward to identify $\eta_{s} = \eta$. We can further relate $\zeta_{s}$ with $\zeta$ and $\chi$. Indeed, from Eqs.~\eqref{eq:coeffs-all} and \eqref{eq:zeta-s}, we have
\begin{equation}
\label{eq:correspn-coeffs}
\begin{aligned}
&
\zeta_{s}
=
\zeta + c_{s}^{2} \chi 
,
\end{aligned}
\end{equation}
where we have used the property $\langle A_{\bf p} E_{\bf p} \rangle_{0} = 0$ to derive this result. Both transport coefficients $\eta_{s}$ and $\zeta_{s}$ are independent of the matching condition employed and, thus, useful quantities to investigate. In particular, if one uses the traditional Landau or Eckart matching conditions, the coefficient $\zeta_{s}$ reduces to the bulk viscosity coefficient.

In the limit $M(T)/T \rightarrow 0$ the asymptotic behavior of $\zeta_{s}$ is
\begin{equation}
\hspace{-1.5cm}
\begin{aligned}
& \frac{\zeta_{s}}{t_{R}(\varepsilon_{0} + P_{0})} 
\sim
\frac{1}{10368} M(T)^{2} \left[ \frac{d}{dT}\left(\frac{M(T)}{T}\right) \right]^{2}(\gamma ^4+10 \gamma ^3+11 \gamma ^2-22 \gamma +120)\Gamma(\gamma+1),
\end{aligned}
\end{equation}
thus it vanishes faster than its matching-dependent counterparts $\zeta$ and $\chi$. Indeed, from Eq. \eqref{eq:correspn-coeffs} and the fact that $c^{2}_{s} = 1/3$, the leading order terms of $\zeta$ and $\chi$ shown in Eqs.~\eqref{eq:coeffs-m=0} cancel each other. Comparison with Eq.~\eqref{eq:cs2-exp-m=0}, shows that $\zeta_{s} \propto (c^{2}_{s} - 1/3)^{2}$, a behavior expected to bulk viscosity, $\zeta$. As $\zeta_{s}$ is a linear combination of $\zeta$ and $\chi$, we recover that $\zeta \propto (c^{2}_{s} - 1/3)^{2}$ only for Landau or Eckart matching conditions, where $\chi \equiv 0$. In the opposite limit, $M(T)/T \rightarrow \infty$, we have
\begin{equation}
\begin{aligned}
&
\zeta_{s} \sim 
 \frac{1}{3} \left( \frac{M(T)}{T} \right)^{\gamma-1}\left\{ 2(1- 3 c_{s}^{2})^{2} +
3 \left[c_{s}^{2} T \frac{d}{dT}\left( \frac{M(T)}{T} \right) + (1- 3 c_{s}^{2}) \right]^{2}
 \right\},
\end{aligned}    
\end{equation}
which behaves with the same power of $(M(T)/T)$ as $\zeta$ and $\chi$ (see Eqs.~\eqref{eq:coeffs-m-inf}). The leading order terms do not cancel because in this regime $c_{s}^{2} \neq 1/3$. 

\begin{figure}[!h]
    \centering
\begin{subfigure}{0.47\textwidth}
  \includegraphics[width=\linewidth]{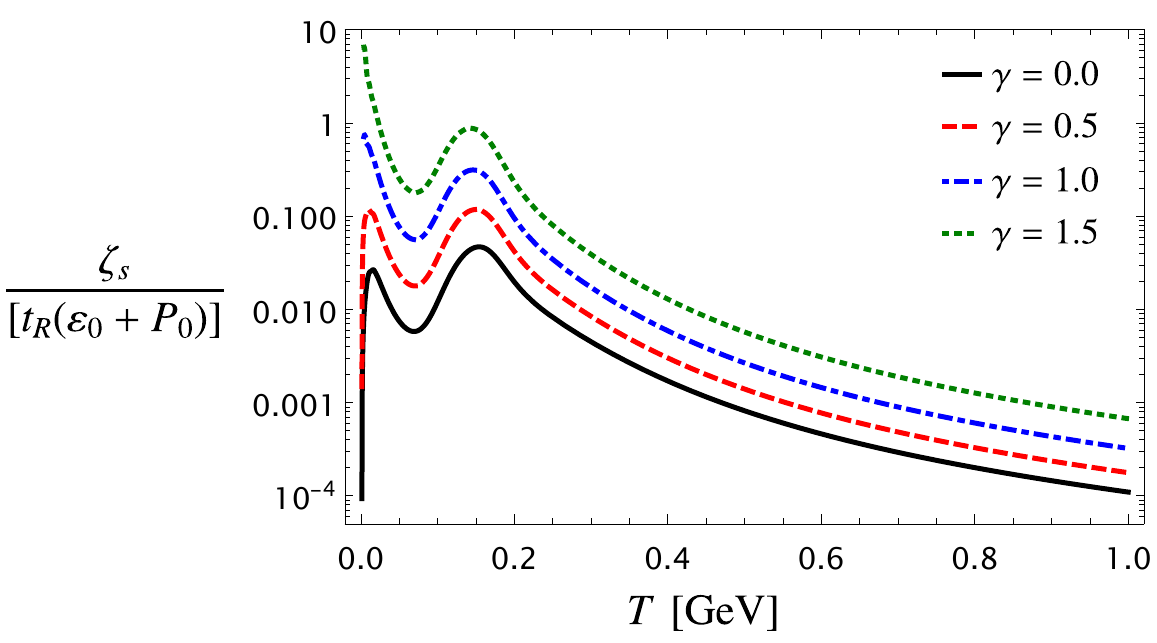}
\end{subfigure}\hfil 
\caption{(Color online) Matching-condition invariant bulk viscosity as a function of temperature in GeV with varying phenomenological parameter $\gamma$.}
\label{fig:bulk-s}
\end{figure}

In Fig.~\ref{fig:bulk-s}, we show how the matching-invariant bulk coefficient $\zeta_{s}$ varies with temperature for various phenomenological parameters $\gamma$. We find a behavior similar to that of Fig.~\ref{fig:eta-zeta-gammas}, where the curves are shifted to larger values as $\gamma$ increases without changing the overall shape for larger temperatures, and it goes from practically vanishing to diverging at low temperatures.

\section{Conclusion}
\label{sec:concl}

In this work, we calculated the transport coefficients of relativistic Navier-Stokes theory arising from an effective kinetic theory
that reproduces the basic thermodynamic properties of QCD \cite{Alqahtani:2015qja} (at zero chemical potential). In this approach, one of the fundamental problems is to determine the dynamical evolution of the effective mean-field that emerges when considering a temperature-dependent mass. Here, we solved this problem by a judicious choice of matching condition, which renders this field solely temperature dependent even when the system is out of equilibrium. In this case, a new transport coefficient appears (in contrast to Navier-Stokes theory in the Landau or Eckart frames \cite{deGroot:80relativistic}) in the scalar sector which describes non-equilibrium corrections to the energy density in the local rest frame. 

Furthermore, we approximated the collision term using the relaxation time approximation \cite{rocha:2021}, which was constructed to be consistent with fundamental properties of the linearized Boltzmann equation that stem from microscopic conservation laws (in contrast to the traditional Anderson-Witting approximation \cite{ANDERSON1974466}). We note that calculations of transport coefficients using the relaxation time approximation from \cite{ANDERSON1974466} could only be consistently calculated when the relaxation time was not energy-dependent.
We found that all transport coefficients are significantly affected by the energy dependence of the relaxation time -- all increasing in magnitude with $\gamma$, where $\tau_R \sim E_{\pp }^\gamma$. The bulk viscosity and the new transport coefficient $\chi$ display qualitatively similar temperature dependencies, but with opposite signs.  

We further investigated the entropy production of this effective kinetic theory and demonstrated that it is positive semi-definite, in agreement with the second law of thermodynamics. Using the entropy production truncated to first order, we identify matching-invariant and non-negative combinations of transport coefficients. In the scalar sector, we find that $\zeta_{s} =
\zeta + c_{s}^{2} \chi$ is independent of the choice of matching condition, which is in agreement with \cite{Kovtun:2019hdm,bemfica:19nonlinear}.

As mentioned above, we restricted our analysis to the fluid-dynamical theory that emerges from the Chapman-Enskog expansion. It would be interesting to go beyond the current study and also compute the transport coefficients that appear in other fluid-dynamical theories, such as the first-order BDNK equations \cite{bemfica:18causality,Kovtun:2019hdm,bemfica:19nonlinear,Hoult:2020eho,bemfica:20general} and/or second-order theories such as  Israel-Stewart theory \cite{Israel:1979wp} and the DNMR equations \cite{Denicol:2012cn}, which have been recently generalized to take into account a general definition of the local equilibrium state in \cite{Speranza:2021bxf} and \cite{rocha21-transient}, respectively. 
  
\section*{Acknowledgements}

\noindent
G.S.R. is financially funded by Conselho Nacional de Desenvolvimento Científico e Tecnológico (CNPq), process No. 142548/2019-7. G.S.D. also acknowledges CNPq as well as Fundação Carlos Chagas Filho de Amparo à Pesquisa do Estado do Rio de Janeiro (FAPERJ), process No. E-26/202.747/2018. M.N.F is supported by the Funda\c c\~ao de Amparo \`a Pesquisa do Estado de S\~ao Paulo (FAPESP) grants 2017/05685-2 and 2020/12795-1. J.N. is partially supported by the U.S.~Department of Energy, Office of Science, Office for Nuclear Physics under Award No. DE-SC0021301.

\appendix

\section{Summation of the $\phi$ components}
\label{sec:sum-coeffs}

In this Appendix, we show how we perform the sum of the series involving the orthogonal polynomials $P^{(\ell)}_{n}$. To that end, we will see that it is not necessary to explicitly construct the set of polynomials, but only use their orthogonality \eqref{eq:polys-orthogonal-2}. Therefore, we assume that the set of polynomials $P^{(\ell)}_{n}(\beta E_{\mathbf{p}})$ ($n=0,1,...$, for a given $\ell$) is a complete basis of the space of functions with a finite norm, given by Eq.~\eqref{eq:inner-prod-l}, so that an arbitrary function of energy can be expanded as 
\begin{equation}
\label{eq:expn-Pnl}
\begin{aligned}
& G(E_{\mathbf{p}}) =  \sum_{n=0}^{\infty} \lambda_{n} P^{(\ell)}_{n}(E_{\mathbf{p}}), 
\end{aligned}
\end{equation}
for a given $\ell$. To find the coefficients $\lambda_{n}$, one resorts to the orthogonality relation \eqref{eq:polys-orthogonal-2}. Thus, integrating both sides with $\frac{E_{\mathbf{p}}}{\tau_{R,p}} \left( \Delta^{\mu \nu} p^{\mu} p^{\nu} \right)^{\ell} P^{(\ell)}_{m}$ one finds
\begin{equation}
\label{eq:coef-expn-Pnl}
\begin{aligned}
& \lambda_{n} = \frac{1}{A_{n}^{(\ell)}} \left\langle \left(\Delta^{\mu \nu} p_{\mu} p_{\nu}\right)^{\ell} \frac{E_{\mathbf{p}}}{\tau_{R,p}} G(E_{\mathbf{p}}) P^{(\ell)}_{n} \right\rangle_{0}.
\end{aligned}    
\end{equation}

As an example, we compute the tensor component of $\phi^{(1)}_{\bf p}$. From Eq.~\eqref{eq:expn-tensorial} and of Eq.~\eqref{eq:soln-Phis-l=2} one has,
\begin{equation}
\label{eq:F2-sum}
\begin{aligned}
& F^{(2)}_{\mathbf{p}} = \beta \sum_{n=0}^{\infty} \frac{1}{A_{n}^{(2)}}  \left\langle  \left(\Delta^{\mu \nu} p_{\mu} p_{\nu}\right)^{2} P^{(2)}_{n} \right\rangle_{0}  P^{(2)}_{n}.
\end{aligned}
\end{equation}
The term inside the sum has the form of Eq.~\eqref{eq:coef-expn-Pnl} for $G(E_{\mathbf{p}}) = \tau_{R}/E_{\mathbf{p}}$ and $\ell = 2$. Using expansion \eqref{eq:expn-Pnl}, one finds that
\begin{equation}
\label{eq:F2-summed}
\begin{aligned}
& F^{(2)}_{\mathbf{p}} = \beta \frac{\tau_{R}}{E_{\mathbf{p}}}.
\end{aligned}
\end{equation}
An analogous procedure is employed for $F^{(0)}$.
Nevertheless, it is important to point out that there is an extra step in the computation, due to the presence of zero-modes in the collision term. Then, from Eqs.~\eqref{eq:expn-tensorial}, \eqref{eq:soln-Phis-l=0} and \eqref{eq:matching-phis}, which stems from the matching fixation procedure 
\begin{equation}
\label{eq:F1-sum}
\begin{aligned}
& F^{(0)}_{\mathbf{p}} = \sum_{n=1}^{\infty} \frac{1}{A_{n}^{(0)}}  \left\langle  A_{\mathbf{p}} P^{(0)}_{n} \right\rangle_{0}  \left( 
\frac{\langle P_{n}^{(0)} \rangle_{0}}{\langle  P_{0}^{(0)} \rangle_{0}} P^{(0)}_{0} - P^{(0)}_{n} \right).
\end{aligned}
\end{equation}
Then, from Eqs.~\eqref{eq:expn-Pnl} and \eqref{eq:coef-expn-Pnl} for $G(E_{\mathbf{p}}) = (\tau_{R}/E_{\mathbf{p}})A_{\mathbf{p}}$, $\ell = 0$ one has,
\begin{equation}
\begin{aligned}
&
\sum_{n=2}^{\infty} \frac{1}{A_{n}^{(2)}}  \left\langle A_{\mathbf{p}} P^{(0)}_{n} \right\rangle_{0} P^{(0)}_{n} = \frac{\tau_{R}}{E_{\mathbf{p}}}A_{\mathbf{p}} - \frac{1}{A^{(0)}_{0}}\left\langle  A_{\mathbf{p}} P^{(0)}_{0} \right\rangle_{0} \beta E_{\bf p} 
\end{aligned}
\end{equation}
At this point, one is reminded that $P^{(0)}_{0} = \beta E_{\bf p}$. Thus, from the fact that  $\left\langle  A_{\mathbf{p}} E_{\bf p} \right\rangle_{0} = 0$,
\begin{equation}
\begin{aligned}
& F^{(0)}_{\mathbf{p}} = - \frac{\tau_{R}}{E_{\mathbf{p}}} A_{\mathbf{p}} 
+
\frac{P^{(0)}_{0}}{\left\langle  P^{(0)}_{0} \right\rangle_{0}} \left\langle \frac{\tau_{R}} {E_{\mathbf{p}}}A_{\mathbf{p}}  \right\rangle_{0}.
\end{aligned}    
\end{equation}

\bibliographystyle{apsrev4-1}
\bibliography{bibliography}

\end{document}